\documentclass[conference]{IEEEtran}

\IEEEoverridecommandlockouts

\usepackage{cite}
\usepackage{amsmath,amssymb,amsfonts}
\usepackage{hyperref}
\usepackage{algorithmic}
\usepackage[dvips]{graphicx}
\usepackage{textcomp}
\usepackage{xcolor}
\def\BibTeX{{\rm B\kern-.05em{\sc i\kern-.025em b}\kern-.08em
    T\kern-.1667em\lower.7ex\hbox{E}\kern-.125emX}}

\newenvironment{Itemize}{\begin{itemize}\setlength{\parsep}{0mm}\setlength{\parskip}{0mm}\setlength{\itemsep}{0.5ex}}{\end{itemize}}

\newcommand{\Bf}[1]{{\bf #1}}
\newcommand{\Cal}[1]{{\cal #1}}

\newcommand{\Funct}[2]{#1\!\left(#2\right)}

\newcommand{\Magn}[1]{\left| #1 \right|}

\newcommand{\Brack}[1]{\left[ #1 \right]}
\newcommand{\Parth}[1]{\left( #1 \right)}

\newcommand{\Floor}[1]{\left\lfloor #1 \right\rfloor}

\newcommand{\Inv}[1]{\frac{1}{#1}}

\newcommand{\Half}{\frac{1}{2}}

\newcommand{\Trns}[1]{{\bf #1}^{\mbox{\sf \tiny T}}}

\newcommand{\Derivf}[2]{\frac{\mathrm{d}{#1}}{\mathrm{d} {#2}}}

\newcommand{\Diag}[1]{\mbox{diag}\!\left[\bf #1 \right]}
\newcommand{\Sign}[1]{\mbox{sign}\!\left( #1 \right)}

\newcommand{\KSum}{\sum_{k=0}^{K-1}}

\newcommand{\DefEq}{\stackrel{\text{\tiny def}}{=}}

\begin{document}

\title{Differentiable bit-rate estimation for neural-based video codec enhancement
\thanks{\textsuperscript{*}Qualcomm AI Research is an initiative of Qualcomm Technologies, Inc.}
}

\author{\IEEEauthorblockN{Amir Said}
\IEEEauthorblockA{\textit{Qualcomm AI Research{\footnotesize \textsuperscript{*}}} \\
San Diego, CA, USA \\
asaid@qti.qualcomm.com}
\and
\IEEEauthorblockN{Manish Kumar Singh}
\IEEEauthorblockA{\textit{Qualcomm AI Research{\footnotesize \textsuperscript{*}}} \\
San Diego, CA, USA \\
masi@qti.qualcomm.com}
\and
\IEEEauthorblockN{Reza Pourreza}
\IEEEauthorblockA{\textit{Qualcomm AI Research{\footnotesize \textsuperscript{*}}} \\
San Diego, CA, USA \\
pourreza@qti.qualcomm.com}
}

\maketitle

\begin{abstract}
Neural networks (NN) can improve standard video compression by pre- and post-processing the encoded video.
For optimal NN training, the standard codec needs to be replaced with a codec proxy that can provide derivatives of
estimated bit-rate and distortion, which are used for gradient back-propagation. Since entropy coding of standard codecs
is designed to take into account non-linear dependencies between transform coefficients, bit-rates cannot be well
approximated with simple per-coefficient estimators.
This paper presents a new approach for bit-rate estimation that is similar to the type employed in training end-to-end
neural codecs, and able to efficiently take into account those statistical dependencies. It is defined from
a mathematical model that provides closed-form formulas for the estimates and their gradients, reducing the
computational complexity. Experimental results demonstrate the method's accuracy in estimating HEVC/H.265 codec bit-rates.
\end{abstract}


\section{Introduction}\label{scIntro}

In consumer devices, video codecs are commonly implemented using custom hardware (ASICs), that provide high performance
but reduce flexibility, since modifications require slow and expensive re-designs and deployment.

\begin{figure}
\centering
\includegraphics[width=76mm]{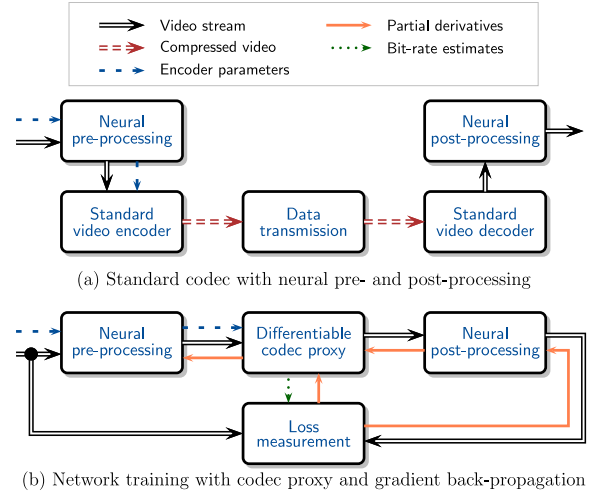}
\caption{\label{fg:TrainSys}Systems for (a) video compression enhancement with neural networks, 
 and (b) corresponding network training.\vspace{-25mm}}
\end{figure}

Codec performance can be improved without ASIC changes by modifying the video before encoding and after decoding, and
the latest trend is to employ neural networks (NN), as shown in Fig.~\ref{fg:TrainSys}(a). Examples of applications
include denoising, artifact removal, resolution changes, etc.~\cite{Ma:20:ivc,Ding:21:avc,Anwar:20:adj}.

Best results are expected with an end-to-end optimization, i.e., NN training that takes into account
codec parameters and performance. A fundamental problem is that NN training is much more effective when it can use
derivatives of performance measurements~\cite{Bottou:10:lml,Goodfellow:16:dpl}, but those are not directly obtainable
from common standard video codec implementations.

The solution is to employ a {\em codec proxy} (e.g.,~\cite{Qiu:21:csn,Luo:20:rda,Guleryuz:21:sic,Chadha:21:dpp}),
that can accurately estimate performance factors and corresponding derivatives, as shown in Fig.~\ref{fg:TrainSys}(b),
enabling NN gradient back-propagation~\cite{Baydin:18:adm,Paszke:17:adp}. In the context of video coding, the loss
function must simultaneously take into account the conflicting objectives of minimizing distortion and bit-rates.

For distortion estimation, the approaches developed for training end-to-end neural codecs (EENCs) provide good
differentiable approximations, and can be used together with  methods to estimate subjective
quality~\cite{Balle:18:vic,Guleryuz:21:sic,Chadha:21:dpp}.

On the other hand, the differentiable bit-rate estimation methods developed for EENCs cannot be directly modified to
the standard codec case, since they are applied to very different types of data, and thus use quite distinct strategies
for optimizing entropy coding.

The problem of bit-rate estimation for video compression is well-known, since it is needed for rate control, which is
fundamental for practical video coding~\cite{Li:14:lrc,Ramanand:17:src}. Several methods use, for example, models that
estimate bit-rates based the quantizer step size $Q$~\cite{Ma:05:qrc,Kwon:07:rcf}.

However, most of those methods are meant to be directly used with the standard codecs, and thus may not be differentiable,
nor suitable to the conditions of  NN training, where it is necessary to obtain estimates at the fine scale of transform
blocks, with per-pixel derivatives.

Methods like the well-known $\rho$-domain rate control~\cite{He:02:rho}, are more suitable for small-scale
estimations, but become less accurate when modified to a differentiable version (cf. eq.~(\ref{eq:LogRate})).

In this paper, we propose a bit-rate estimation designed to work with the data used by the standard codecs, but that
achieves higher accuracy by using a form of data modeling that is similar to that used in training neural codecs.
We show that, since it is based on a mathematical formulation, it is possible to derive closed-form equations for the 
estimate and corresponding derivatives, enabling more efficient computations and faster NN training.

In the next section we discuss why advanced entropy coding makes accurate bit-rate estimation difficult, and in 
Section~\ref{scModel} we present the proposed statistical model, and its similarity to what is used in EENCs.
Section~\ref{scEstim} presents the formulas and computation methods, and the experimental results are shown and
discussed in Section~\ref{scResults}.

\section{Entropy coding in video codecs}\label{scEntropy}

\begin{figure}
\centering
\includegraphics[width=82mm]{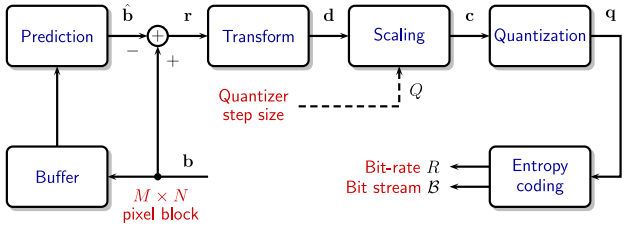}
\caption{\label{fg:Encoder}Simplified diagram of the hybrid video encoding used in standard codecs.\vspace{-15mm}}
\end{figure}

Fig.~\ref{fg:Encoder} shows a diagram of the hybrid coding architecture used by standard video codecs, and
introduces the notation used in this document. At a given encoding stage a block of $M \times N$ pixel is
predicted, an orthogonal transform is applied to the vector $\Bf{r}$ of prediction residuals, the resulting vector $\Bf{d}$
is divided by quantizer step size $Q$ to obtain the scaled coefficients $\Bf{c}$, that are finally quantized and
entropy-coded. 

In this notation all vectors have dimension $K=MN$, and to maintain consistency with signal processing notation,
all vector and matrix indexes start at zero.

One simple way to estimate bit-rates is to sum per-coefficient estimates. For
example, the differentiable approximation to $\rho$-domain estimation used in~\cite{Guleryuz:21:sic} is
\begin{equation}
 \hat{R}_d(\Bf{c}) = \mu \KSum \log_2(1 + |c_k|), \label{eq:LogRate}
\end{equation}
where $\mu$ is a factor obtained from JPEG bit-rates.

The main problem with this approach is that transform coefficients are not coded separately, since they are not
statistically independent. In fact, exploiting magnitude dependencies yields large compression gains, and
motivates the adoption of quite complex forms of entropy coding~\cite{Richardson:10:avc,Sze:12:htc,Sze:14:ech,Wien:15:HEV}.

For this reason many implementations use the standard's actual entropy encoding method for bit-rate estimation.
While this approach is the most precise, there are many practical problems in adapting it to create a differentiable
version.

Differentiable approximations have been developed for solving the problem that quantization derivatives are zero
nearly everywhere~\cite{Shin:17:jpg,Luo:20:rda,Guleryuz:21:sic}. While they are quite useful for distortion estimates,
it is much harder to employ them in complex entropy coding processes based on quantized values.

For example, coefficients quantized to zero are commonly coded together, or by signaling the position of the
last nonzero element. Nonzero values are binarized and can be coded with a variable number of passes, using
different coding contexts per binary symbol~\cite{Sze:12:htc,Sze:14:ech}.

Those difficulties motivate searching for better bit-rate estimation methods, based on the same statistical properties,
but using a different methodology.

\section{Model-based estimation}\label{scModel}

\begin{figure}
\centering
\includegraphics[width=60mm]{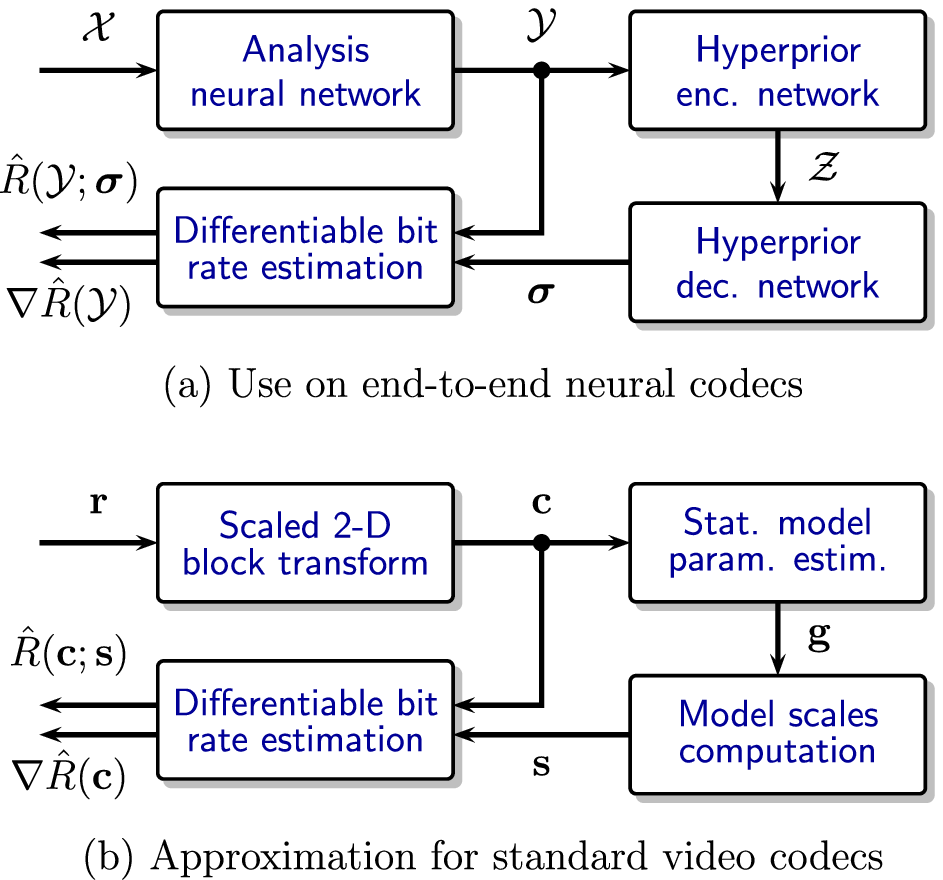}
\caption{\label{fg:EstComp}Comparison of systems for bit-rate estimation based on statistical models.\vspace{-15mm}}
\end{figure}

Fig.~\ref{fg:EstComp}(a) shows the basic structure used for training an EENC with a hyper-prior architecture~\cite{Balle:18:vic}.
It is based on establishing an statistical model of the non-linear transform elements (to be entropy coded
after quantization), defined by the distribution type (commonly Gaussian), and array {\boldmath $\sigma$} with standard deviations.

This approach can naturally incorporate the statistical dependencies among many data elements, translated into the
variations on standard deviation values, and experimental results have shown that it yields quite accurate bit-rate estimates.

For those reasons, we propose a similar approach, shown in Fig.~\ref{fg:EstComp}(b), with the following differences
\begin{Itemize}
\item It is applied to coefficients of an orthogonal transform, like discrete cosine or sine, used by the 
 standard codec.
\item Transform coefficients are assumed to be zero-mean random variables with Laplace distributions, and the standard
 deviations are defined by a model with a few parameters in vector $\Bf{g}$.
\item The statistical model is based on the empiric observation that the variance of transform coefficients
tends to decrease exponential with frequency~\cite{Said:08:epd}, with decrease rate depending on orientation of pixel
patterns
\item For each block, the maximum-likelihood (ML) parameters $\Bf{g}^*$ are computed, using all coefficient values, and
 the model and $\Bf{g}^*$ are used to estimate bit-rate and gradient.
\end{Itemize}

\section{Practical implementation}\label{scEstim}

\begin{figure}
\centering
\includegraphics[width=82mm]{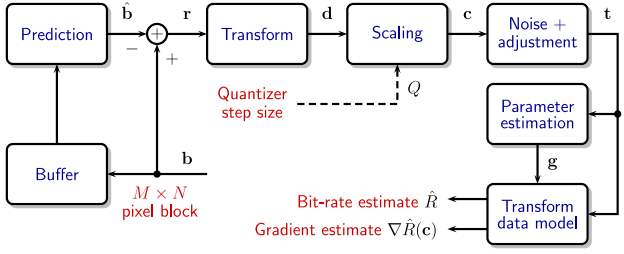}
\caption{\label{fg:BitRate}Proposed system for bit-rate estimation.\vspace{-15mm}}
\end{figure}

There are many practical details in the implementation of a codec proxy that are outside the scope of this paper.
As shown in Fig.~\ref{fg:BitRate}, we assume the main prediction parts of a hybrid encoder, shown in Fig.~\ref{fg:Encoder},
are approximated. Similarly, the choice of $Q$ can be fixed or change randomly~\cite{Chadha:21:dpp}, depending
on training objectives.

The main strategy is what was outlined in the previous section, and in this section we present approximations
needed for a practical implementation, plus some heuristics that were shown to improve accuracy and numerical stability.

To use indexes that are related to two-dimensional transforms, vectors $\Bf{m}$ and $\Bf{n}$ are defined by
\begin{equation}
  m_k \DefEq \Floor{ k / N }, \quad n_k \DefEq k - N \, m_k,
\end{equation}
and using $\Bf{1}$ to represent the all-ones vector, we define the following $K \times 3$ matrix
\begin{equation}
  \Bf{A} \DefEq \begin{bmatrix} \Bf{1} & \Bf{m} & \Bf{n} \end{bmatrix}.
\end{equation}

\subsection{Coefficient adjustment}

\begin{figure}
\centering
\includegraphics[width=70mm]{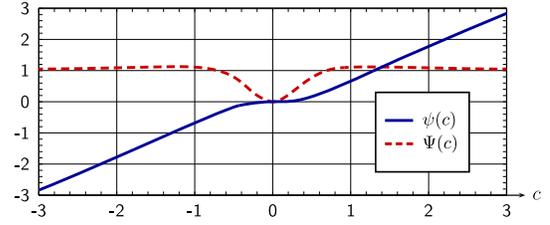}
\caption{\label{fg:Adjust}Function used for transform coefficient adjustments, and its derivative.\vspace{-20mm}}
\end{figure}

To approximate the common dead-zone~\cite{Sullivan:05:odz} or R-D optimized~\cite{Karczewicz:09:rdo}
quantization, transform coefficients are first ``adjusted'' to reduce small magnitudes, using the function shown
in Fig.~\ref{fg:Adjust}, together with its derivative
\begin{eqnarray}
 \psi(c) & \DefEq & \frac{c^3}{c^2 + \tau}, \label{eq:AdjFunc} \\
 \Psi(c) & \DefEq & \Derivf{\psi(c)}{c} = 1 + \frac{\tau \Parth{ c^2 - \tau }}{\Parth{ c^2 + \tau }^2}. \nonumber
\end{eqnarray}

\subsection{Noise addition}

Uniform noise is added to avoid numerical instability when all coefficients are zero or very small. 
Given an array of uniformly distributed random variables $\eta_k \sim U(-\epsilon, \epsilon)$, we define vectors
$\Bf{t}$ and $\Bf{w}$, used for estimation
\begin{equation}
 t_k \DefEq \psi(c_k), \quad w_k \DefEq \Magn{ t_k + \eta_k }. \label{eq:Adjust}
\end{equation}

\subsection{Probability distribution model}

For estimating model parameters, it is assumed that elements of vector $\Bf{w}$ have exponential
probability distribution, and their standard deviation decay exponentially with frequency according to 3-dimensional
parameter vector $\Bf{g}$, as
\begin{equation}
  \sigma_k(\Bf{g}) = \Funct{\exp}{-[g_0 + m_k g_1 + n_k g_2]}.
\end{equation}

To simplify notation we define the vector with standard deviation reciprocals
\begin{equation}
  s_k(\Bf{g}) \DefEq 1 / \sigma_k(\Bf{g}) = \Funct{\exp}{g_0 + m_k g_1 + n_k g_2},
\end{equation}
to obtain the probability distribution functions
\begin{equation}
 f(w_k; s_k(\Bf{g})) = s_k(\Bf{g}) e^{-s_k(\Bf{g}) w_k}. \label{eq:pdf}
\end{equation}

\subsection{Maximum-likelihood parameters}

The likelihood function defined by distributions in~(\ref{eq:pdf}) is
\begin{equation}
 \Cal{L}(\Bf{g}) = \prod_{k=0}^{K-1} f(w_k; s_k(\Bf{g})),
\end{equation}
and the negative of the log-likelihood is
\begin{equation}
 L(\Bf{g}) \DefEq -\Funct{\log}{\Cal{L}(\Bf{g})} = \Trns{w} \Bf{s}(\Bf{g}) - \Trns{1} \Bf{A} \Bf{g}.
\end{equation}

Using $\circ$ to represent per-element vector multiplications, and considering that the gradient
\begin{equation}
 \nabla L(\Bf{g}) = \Trns{A} \Brack{ \Bf{w} \circ \Bf{s}(\Bf{g}) - \Bf{1} },
\end{equation}
and $3\times 3$ symmetric Hessian matrix
\begin{equation}
 \Bf{H}(\Bf{g}) = \Trns{A} \Diag{\Bf{w} \circ \Bf{s}(\Bf{g})} \Bf{A},
\end{equation}
are easy to compute, the maximum-likelihood solution can be found, for example, applying Newton's iterations
\begin{equation}
 \Bf{g} \leftarrow \Bf{g} - \Brack{ \Bf{H}(\Bf{g}) }^{-1} \nabla L(\Bf{g}), \label{eq:NewtIter}
\end{equation}
which should, with proper implementation~\cite{Press:07:asc}, converge to optimal solution $\Bf{g^*}$. 

Note that $3 \times 3$ symmetric matrix inversions, or a form of Cholesky decompositions, can be easily computed.

\subsection{Bit-rate estimation}

With the maximum-likelihood probability distribution parameters
$\Bf{s}^* \DefEq \Bf{s}(\Bf{g}^*)$,
we can use the technique developed for end-to-end neural codecs~\cite{Balle:18:vic} to obtain differentiable estimates of
the bit-rates, assuming that adjusted parameters $t_k$ have Laplace probability distribution, with the
cumulative distribution function in the form
\begin{equation}
 F(t; s) = \begin{cases}
   \Half e^{s t}, & t < 0, \\ 
   1 - \Half e^{-s t}, & t \geq 0.
  \end{cases}
\end{equation}
and parameters $s_k^*$. This is not mathematically exact, due to noise addition in~(\ref{eq:Adjust}),
but is a convenient approximation.

The differentiable estimated probability of the quantized transform coefficient is given by
\begin{equation}
 p_k = F(t_k + 1/2; s_k^*) - F(t_k - 1/2; s_k^*),
\end{equation}
and the bit-rate is estimated from the entropy equation
\begin{equation}
 \hat{R}(\Bf{c}) = - \frac{\alpha}{K} \KSum \Funct{\log_2}{p_k}, \label{eq:Rate}
\end{equation}
where multiplicative factor $\alpha$ is added for calibration, similarly to parameter $\mu$ in eq.~(\ref{eq:LogRate}).

Note that during training, bit-rates must be multiplied by a factor before being added to distortion. 
This factor depends on the training objectives, and optimal values can only be determined through validation tests.

For example, experimental tests can show that a certain value of $\alpha$ can be best for H.264/AVC,
and another value for H.265/HEVC. The main objective is to have consistency in the estimates, so that
design choices are correctly based on video characteristics.

\subsection{Partial derivative computations}

Since all stages in the derivation of~(\ref{eq:Rate}) are differentiable, gradient $\nabla \hat{R}(\Bf{c})$ can
be effectively and easily computed using automatic differentiation~\cite{Baydin:18:adm,Paszke:17:adp}.

However, the use of Newton iterations to determine $\Bf{g}^*$ requires creating sequences of vectors
$\Bf{g}^{(0)}, \Bf{g}^{(1)}, \ldots$, which adds extra computations during gradient back-propagation.

Those computations can be eliminated by exploiting the mathematical properties of the model's formulation.
It can be shown that, defining functions
\begin{eqnarray}
 \gamma(k, \delta) & \DefEq & \frac{\alpha s_k^* \Funct{\exp}{-s_k^* |t_k + \delta|}}{2 \ln(2) K p_k}, \\
 \phi(k, \delta) & \DefEq & (t_k + \delta) \, \gamma(k,\delta), \nonumber
\end{eqnarray}
and vectors
\begin{eqnarray}
 u_k & \DefEq & \gamma(k, 1/2) - \gamma(k, -1/2), \\
 v_k & \DefEq & \phi(k, 1/2) - \phi(k, -1/2), \nonumber \\
 y_k & \DefEq & \Psi(c_k), \nonumber \\
 z_k & \DefEq & \Sign{t_k + \eta_k} s_k^*, \nonumber
\end{eqnarray}
the bit-rate gradient can be computed directly and more efficiently using the equation
\begin{equation}
 \nabla \hat{R}(\Bf{c}) = \Bf{y} \circ \Brack{ \Bf{z} \circ \Parth{ \Bf{A} \Brack{ \Bf{H}(\Bf{g}^*) }^{-1} \Trns{A} \Bf{v} } - \Bf{u} }.
\end{equation}

\begin{figure}
\centering
\includegraphics[width=56mm]{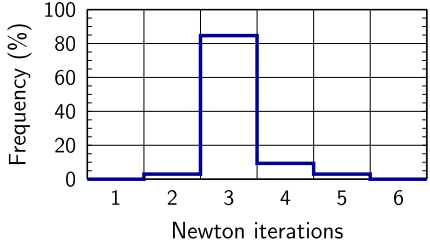}
\caption{\label{fg:NumIter}Distribution of number of iterations to achieve high precision.\vspace{-8mm}}
\end{figure}

Note that, even though each term of $\nabla \hat{R}(\Bf{c})$ depends on all elements of $\Bf{c}$, the efficient 
computation of intermediate results allows the computation to be done with $\mathrm{O}(MN)$ instead of
$\mathrm{O}(M^2 N^2)$ complexity, and it is easy to optimize the implementation and parallelize vector and matrix operations.

\section{Experimental results}\label{scResults}

The proposed method was tested to estimate bit-rates of the H.265/HEVC codec. HM~16.20 reference implementation~\cite{HEVC:HM}
was modified to output the DCT of block residuals, and the resulting estimates were compared to the number of bits actually used 
for each frame.
\begin{figure}
\centering
\includegraphics[width=84mm]{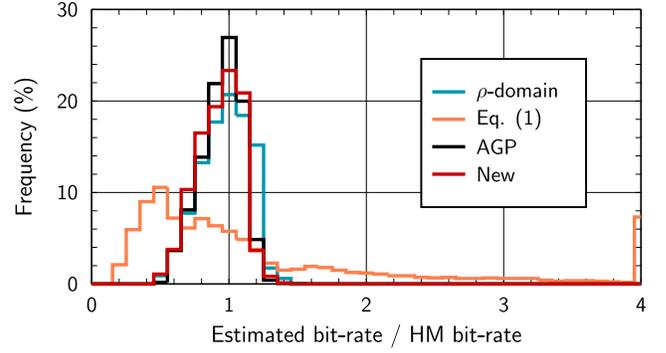}
\caption{\label{fg:HistAll}Histograms of ratios between estimated and HM bit-rates from different differentiable and
non-differentiable methods.}
\end{figure}

Experiments were performed using a low-delay-P configuration, on test videos of VVC standardization, classes A1, A2,
B, and~E, with QP = 22, 27, 32, and 37. For consistency, all 14~videos were converted to $1280 \times 720$ HD resolution,
250~frames per video, for a total of 14,000 frames tested. 

The estimation used only the luma component in all tested methods. Due to lack of space, only $8\times 8$ block results
are reported here, and that was the forced transform size.

The method was implemented using $\tau=0.4$ in eq.~(\ref{eq:AdjFunc}), and $\eta=0.05$ was used for uniform noise
generation (significantly smaller than used for EENC training). 

In all tests the initial solution was $g_1=g_2=0.05$ and
\begin{equation}
 g_0 = - \Funct{\ln}{\Inv{K} \KSum w_k \, e^{g_1 m_k + g_2 n_k}}. \label{eq:NewtInit}
\end{equation}
Fig.~\ref{fg:NumIter} shows the observed distribution of the number of Newton iterations~(\ref{eq:NewtIter}),
using this initialization. It can be seen that, in the majority of cases, sufficiently high
precision is achieved in only 3~iterations.

Fig.~\ref{fg:HistAll} shows histograms of ratios between bit-rates from some estimation methods and actual HM bit-rates.
In this type of figure an ideal estimator would have 100\% of the ratios around one. All tested methods used calibration
coefficients optimized on the four QP values, to measure their accuracy over a wide range of bit rates.

\begin{table}
\renewcommand{\arraystretch}{1.25}
\centering
\caption{\label{tb:EstVar}Standard deviation of the estimation ratios, according to HM QP values.}
\begin{tabular}{|c|c|c|c|c|c|} \hline
 \bf QP & \bf Avrg. &\multicolumn{4}{|c|}{\bf Estimation method} \\ \cline{3-6}
    & \bf bit-rate & $\rho$-domain & AGP & Eq. (1) & Proposed \\ \hline \hline
 22 & 0.257 & 0.158 & 0.121 & 0.817 & 0.126 \\
 27 & 0.120 & 0.131 & 0.125 & 0.708 & 0.129 \\
 32 & 0.059 & 0.149 & 0.131 & 0.678 & 0.147 \\
 37 & 0.030 & 0.252 & 0.222 & 0.709 & 0.243 \\ \hline
All &  ---  & 0.179 & 0.156 & 0.730 & 0.168 \\ \hline
\end{tabular}
\end{table}

The best results are obtained using the AGP method~\cite{Said:97:lcw} for context-based entropy coding (a simpler
entropy coding method), and somewhat worse results are obtained using using the $\rho$-domain estimator~\cite{He:02:rho}.
However, those are non-differentiable estimators.

The proposed method yields accuracy between AGP and $\rho$-domain, while being differentiable. The performance of the
differentiable estimator of eq.~(\ref{eq:LogRate})~\cite{Guleryuz:21:sic}, on the other hand, is significantly less
accurate (note that about 8\% of the ratios are actually off-scale, beyond 4), indicating the shortcomings of all forms of 
per-coefficient estimations.

Bit-rate estimation is easier in high-rate settings, and this can be observed by measuring the standard deviation of
ratios measured / actual bit-rates, for different QP values, as shown in Table~\ref{tb:EstVar}. We can observe
that, as the average bit rate varies by about one order of magnitude, the general pattern is the same observed in
Fig.~\ref{fg:HistAll}.

AGP provides the most accurate and consistent results in all bit rates, closely followed by the proposed method, with
accuracy decreasing mostly for lower rates (QP = 37).  The $\rho$-domain estimator is slightly less consistent, while
the per-coefficient estimator has standard deviations that are significantly larger in all tests.

\section{Conclusions}\label{scConclusions}

The experimental results confirm the advantages of using the approach proposed in Section~\ref{scModel}
\begin{Itemize}
\item Bit-rate estimations are much more precise when they, like entropy coding methods, take into account
the statistical dependencies between magnitudes of transform coefficients.
\item Employing a statistical model, with magnitude dependencies defined by distribution of standard deviations
(as used when training end-to-end neural codecs), greatly increase estimation accuracy.
\item Using a proper mathematical formulation allows for direct computations of estimates and their derivatives,
and reduction of computational complexity.
\end{Itemize}


\bibliographystyle{IEEEbib}

\end{document}